\documentclass[sigplan,noacm]{acmart}
\acmSubmissionID{3600}
\renewcommand\footnotetextcopyrightpermission[1]{}
\usepackage[]{hyperref}
\usepackage{xspace}
\usepackage{subcaption}
\usepackage{booktabs}
\usepackage{multirow}
\usepackage[linesnumbered,ruled,vlined,noend]{algorithm2e}
\usepackage{soul}

\newcommand{\sys}[0]{{T-MAN}\xspace}

\newcommand{\ting}[1]{\textcolor{brown}{}}

\newcommand{\remove}[1]{}

\newcommand{\code}[1]{{#1}}

\newcommand\blfootnote[1]{%
\begingroup
\renewcommand\thefootnote{}\footnote{#1}%
\addtocounter{footnote}{-1}%
\endgroup
}

\hypersetup{
    colorlinks=true,  
    urlcolor=blue,    
    linkcolor=blue,   
    citecolor=blue    
}

\begin{document}


\title{\sys{}: Enabling End-to-End Low-Bit LLM Inference on NPUs via Unified Table Lookup} 

\date{}  
\author{  \textbf{
    Jianyu Wei$^{1,*}$ \hspace{1mm} 
    Qingtao Li$^{2,*}$ \hspace{1mm} 
    Shijie Cao$^{2}$ \hspace{1mm}   
    Lingxiao Ma$^{2}$ \hspace{1mm}
    Zixu Hao$^{3}$ \hspace{1mm}
    Yanyong Zhang$^{1}$ \hspace{1mm}
    Xiaoyan Hu$^{4}$ \hspace{1mm}
    Ting Cao$^{5,\dagger}$ \hspace{1mm} 
    }
    \\  
    \textsuperscript{1}University of Science and Technology of China \hspace{1em}  
    \textsuperscript{2}Microsoft Research \hspace{1em} 
    \textsuperscript{3}Tsinghua University \hspace{1em} 
    \textsuperscript{4}Microsoft \hspace{1em} 
    \textsuperscript{5}Institute for AI Industry Research, Tsinghua University \hspace{1em} 
}









\hypersetup{
    colorlinks=true,  
    urlcolor=blue,    
    linkcolor=blue,   
    citecolor=blue    
}

\begin{abstract}
Large language models (LLMs) are increasingly deployed on customer devices. 
To support them, current devices are adopting SoCs (System on Chip) with NPUs (Neural Processing Unit) installed. Although high performance is expected, LLM inference on NPUs is slower than its CPU counterpart. The reason is that NPUs have poor performance on computations other than GEMM, like dequantization. Current works either disaggregate prefill on the NPUs and decoding on the CPUs, or put both on the NPUs but with an accuracy loss.

To solve this issue, based on the insight that low-bit can enable target computation encoded within an acceptably sized table, we propose table lookup to subsume hardware operations otherwise unsupported. To realize this, we overcome the conflicting hardware behavior of prefill and decoding to design a unified table layout and tiling through (1) fused two-level table-based dequantization and (2) concurrency-hierarchy-guided tiling. Based on that, we implement the prefill phase by three-stage pipeline and map the table-lookup-based decoding to NPU's vector units. Results show $1.4\times$ and $3.1\times$ speedup for prefill and decoding respectively, and 84\% energy savings compared to the baseline NPU methods. The code is available at \href{https://github.com/microsoft/T-MAC/tree/main/t-man}{https://github.com/microsoft/T-MAC/tree/main/t-man}.  
\end{abstract}

\pagestyle{plain}
\maketitle 

\blfootnote{* Jianyu Wei and Qingtao Li have equal contributions to this article.}
\blfootnote{$\dagger$ Corresponding to Ting Cao <tingcao@mail.tsinghua.edu.cn>.}

\section{Introduction}


Large language models (LLMs) are becoming an integral part of system services on customer devices, to provide unprecedented and always-available AI experiences. Major customer product companies have already deployed LLMs directly in their flagship products, such as Apple Intelligence on iPhone and Mac~\cite{apple,zhou2025appleintelligencefoundationlanguage}, Google Android on Samsung Galaxy~\cite{gemma}, and Microsoft Windows on Copilot PC~\cite{phisilica}. Due to strict memory constraints, these deployments universally rely on low-bit weight and activation representations. For example, Apple’s foundation models use 2–4 bit~\cite{apple}, Windows Copilot adopts the 4-bit Phi-Silica model~\cite{phisilica}, and Google’s Gemma for mobile runs in 4 bit~\cite{gemma}. More aggressive methods employing $2$-bit and even $1$-bit precision continue to emerge~\cite{du2024bitdistiller,wang2023bitnet}.

Hardware-wise, consumer devices are increasingly equipped with SoCs featuring dedicated NPUs~\cite{npulist} for efficient AI inference, particularly in the LLM era. For example, the Qualcomm Snapdragon SoC with NPU has been used by dozens of different device modules, including Microsoft Surface, Samsung Galaxy, Honor Magic, etc., to support LLMs on devices~\cite{snapdragonlist}. Modern NPUs are designed to have the utmost performance and efficiency to run dense GEMMs. Qualcomm NPU is claimed to have 45\,TOPS peak performance, 150$\times$ of its CPU and 9$\times$ of its GPU counterparts~\cite{xelite}.

However, despite their impressive peak computation performance, low-bit decoding on NPUs is often slower than on CPUs as other work measured~\cite{t-mac}. 
The fundamental issue is that the utmost efficiency of NPU comes at the cost of flexibility. It is specifically optimized for dense GEMM with specific numerical representation (e.g., INT8) and quantization granularity (e.g., channel-wise), but very poor performance for others. For the diverse low-bit LLM format, when it does not match the native NPU GEMM format, the weights must be dequantized online. This dequantization involves element-wise float-point operations, an operation pattern for which NPUs are not optimized.

Current practice addresses this mismatch in two unsatisfactory ways:\ting{we need to survey and add new NPU works} (1) \textit{To align LLM quantization with hardware supported format~\cite{qnnsdk}, but with an accuracy loss.} It avoids runtime dequantization, but our evaluation shows that adopting hardware per-channel quantization causes a 1.45$\times$ worse perplexity, compared to standard quantization methods. (2) \textit{To run the prefill phase on the NPU and decoding phase on the CPUs~\cite{llm.npu}, but with energy overhead.} The prefill is GEMM-dominant, which allows the NPU’s GEMM throughput to offset the dequantization cost. The decoding is memory-intensive GEMV-dominant, so 
the dequantization overhead prevails. However, decoding on CPU consumes 79\% more energy and competes with other applications co-run on the customer devices.\ting{Also mention the double memory cost issue here? Emphasize memory is important.}

\begin{figure}[t]
\centering
\includegraphics[width=0.9\linewidth]{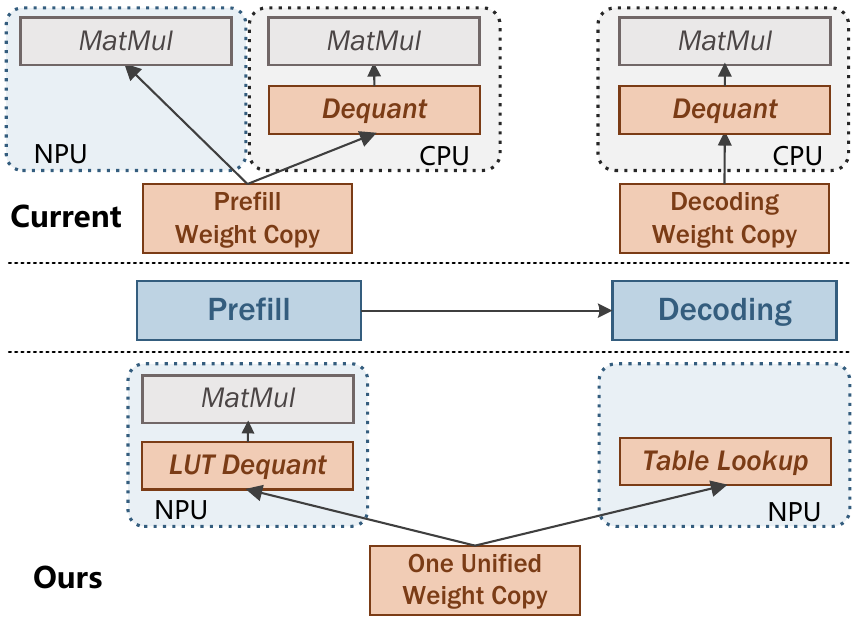} 
\caption{T-MAN versus current practice. To maintain accuracy, current practice offloads decoding phase to CPU and store two weight copies for NPU and CPU respectively. T-MAN leverages table lookup to enable both prefill and decoding on the NPU and only keep one weight copy. \ting{This figure needs slightly improved. Mark different format for Weight-copy; mark power cost above; mark two phases: prefill and decoding. }} 
\label{fig:high-level-idea}
\end{figure}

This paper aims to achieve the most efficient LLM inference on customer device SoCs, by enabling end-to-end inference on the NPU with superior performance for both prefill and decoding without sacrificing accuracy. Our \textit{key insight} is that if the target computation can be encoded within an acceptably sized table, a lookup-table (LUT) mechanism can subsume hardware operations otherwise unsupported. Low-bit quantization exponentially shrinks the space of possible values, making the requisite tables small enough to reside in on-chip memory.  
Based on this insight, this paper explores table lookup to bridge the gap between flexible quantization and NPU-specialized execution by eliminating the runtime dequantization cost, as illustrated in Fig.~\ref{fig:high-level-idea}. 

Though the idea is clear, directly applying it on NPUs faces two major challenges.\ting{emphasize that although there are many table lookup works, current table lookup works like T-MAC cannot solve the issue.} 
(1) \textbf{The conflict of table data layout between prefill and decoding}. In prefill phase, it is necessary to dequantize each low-bit weight to the hardware-specialized GEMM, which requires the bit-parallel table data layout. While for decoding, to eliminate dequantization needs bit-serial data layout to control the table size. Naïvely supporting both doubles on-chip storage. (2) \textbf{The conflict of table lookup tiling of prefill and decoding}. The matrix cores for prefill and vector cores for decoding have different instruction widths and computation patterns, leading to different loop orders and tile sizes. This prevents contiguous storage access across both stages. 

We propose two techniques accordingly to realize a unified table layout and tiling scheme for both prefill and decoding: (1) \textbf{Two-level table lookup} to unify the bit-parallel and bit-serial data layout. (2) \textbf{Concurrency-hierarchy-guided tiling scheme} to unify the tiling of table lookup between prefill and decoding. Based on the unified data layout and tiling, we realize the prefill phase by a DMA-vector-matrix three-phase pipeline to parallely conduct dequantization and matrix multiplication, and realize decoding phase by mapping the LUT-based decoding on vector cores of NPU. 

\ting{Emphasize lossless. Compared to QNN, accuracy better. }
We implement \sys{} based on ExecuTorch~\cite{executorch} and evaluate it on mobile devices with Llama 3, Qwen 3 and BitNet with four different bit formats. Compared to the SOTA LLM inference methods on device (including llm.npu~\cite{llm.npu}, T-MAC~\cite{t-mac}, and llama.cpp~\cite{llama.cpp}, bitnet.cpp~\cite{bitnet.cpp}), we achieve up to $1.4\times$ speedup for prefill and $3.1\times$ for decoding, and up to $84\%$ energy saving compared to the current SOTA NPU solution.

Our contributions can be summarized as follows: 
\begin{itemize}
    \item We realize the first end-to-end LLM inference on NPUs with SOTA speed and energy efficiency compared to other on-device LLM inference systems. 
    \item We propose the novel fused two-level table lookup method to unify the table layout and the concurrency-hierarchy-guided tiling scheme to unify the tiling for both prefill and decoding.
    \item Based on the unified data layout and tiling, we realize the DMA-Vector-Matrix pipeline for prefill and map the LUT-based GEMV for decoding. 
\end{itemize}




\section{Background and Motivation}

\subsection{LLMs on Edge and Low-bit Quantization}
LLMs are increasingly integrated as a foundation system service to deploy on customer devices, such as Gemini on Android, Apple Intelligence on iOS, and Phi in Windows Copilot, to provide real-time, always-on, and privacy-preserving AI services.  


LLM inference consists of two distinct phases: (1) \textit{prefill}, which processes the input prompt in parallel and is highly compute-intensive, and (2) \textit{decoding}, which generates tokens autoregressively and is dominated by memory-bound operations, particularly in the common single-batch scenarios on consumer devices. The challenge becomes more pronounced with emerging reasoning-oriented LLMs~\cite{deepseekai2025deepseekr1incentivizingreasoningcapability,qwen3,openai2024openaio1card}, which often produce thousands of tokens per response.

To reduce the heavy memory bandwidth demands of decoding, low-bit quantization has become indispensable in real-world deployments. Notable examples include Google’s 4-bit Gemma\cite{gemma}, Microsoft’s 4-bit Phi\cite{phisilica} and 1.58-bit BitNet\cite{ma2025bitnetb1582b4ttechnical}, and Apple’s 2–bit foundation models\cite{zhou2025appleintelligencefoundationlanguage}. Meanwhile, new quantization algorithms with diverse bit-widths and granularities continue to emerge at a rapid pace.




\subsection{Low-Bit LLM Inference}
\label{sec:low-bit-approach}

There are diverse quantization formats used by different algorithms for various application scenarios~\cite{llama.cpp,frantar2022gptq,lin2023awq,xiao2023smoothquant,dettmers2022llmint88bitmatrixmultiplication,mxfp4,du2024bitdistiller,xu2024onebit}. The formats can differ in bit widths (e.g., 8-, 4-, 2-, and 1-bit), numerical representation (e.g., FP and INT) and quantization granularity (e.g., group-wise with sizes of 32-, 64-, or 128-, as well as channel- and tensor-wise). Weights and activations normally employ different formats. No single quantization format has emerged as dominant. Contradictorily, hardware designs are impossible to simultaneously support all these formats, due to the limited chip area.

To bridge the gap between the algorithm flexibility and hardware, there are two main methods used currently.                                                                      
\paragraph{Dequantization-based low-bit inference} Dequantization converts the quantized values back into a higher precision to match the hardware supported formats.  The conversion is performed element-wise as: $x_{dequantized} = (x_{quantized} - zero\_point)\times scaling\_factor$. The resulting dequantized values can then be processed directly by existing hardware units. This approach has been widely adopted for low-bit LLM deployment~\cite{bitblas,llm.npu,llama.cpp,vllm}, as it leverages mature hardware for GEMM. However, it introduces extra computation overhead due to the dequantization step.\ting{emphasize the high-cost of dequant again.}

\begin{figure}[t]
\centering
\includegraphics[width=\linewidth]{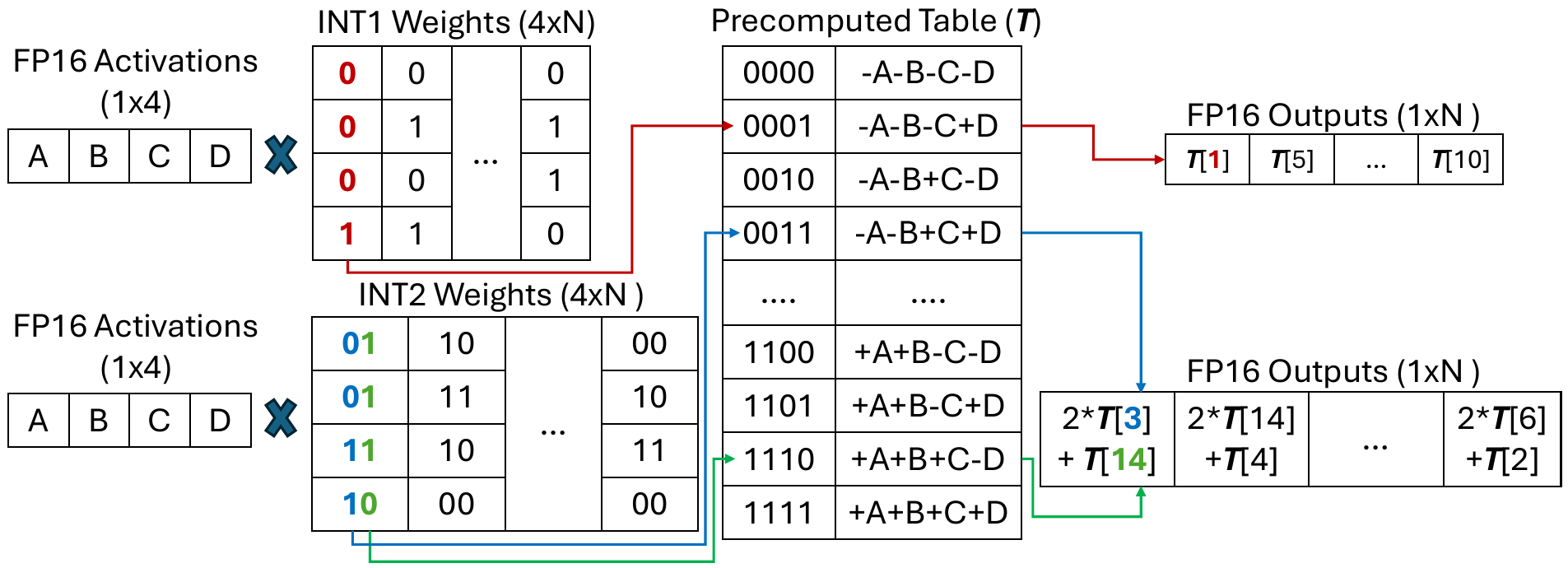} 
\caption{Bit-serial table lookup to implement GEMM. Weight is decomposed into one-bit matrices, serving as the indices to precomputed results stored in tables to look up.\ting{explain or mark the lines as table lookup.}}
\label{fig:lut-gemm}
\end{figure}

\paragraph{Bit-serial table lookup} To eliminate the dequantization, recent works, such as LUT-GEMM~\cite{park2023lutgemm}, T-MAC~\cite{t-mac}, and LUT tensor core~\cite{luttensorcore}, propose a unified computation paradigm for diverse low-bit formats, i.e., bit-serial table lookup. As Fig.~\ref{fig:lut-gemm} shows,  the key idea is based on the linear equivalent transformation of matrix multiplication: the multiplication of two matrices can be transformed into multiplying one matrix by each one-bit decomposition of the other, followed by shifting and accumulating the partial results. Therefore, the matrix multiplication of diverse quantization formats is reduced to unified operations between the activation matrix and the one-bit weight matrices.

Though current hardware lacks native one-bit computation support, the limited value set (1 and 0) of one-bit weights enables implementation via lookup tables. Each vector pattern of the one-bit weight matrix serves as an index to precomputed results stored in tables, to read the vector results directly. This approach makes inference independent of quantization format and avoids dequantization. The downside, however, is that current NPUs provide much lower-performance support for table lookup instructions compared to GEMMs. It is challenging to achieve practical speedup. ~\ting{? This is not right here. We did not mention lower-performance of LUT on NPU as a challenge.}
   
Both approaches have clear trade-offs. Dequantization-based inference fully exploits specialized hardware units but incurs extra overhead from the conversion step. Bit-wise table lookup avoids dequantization and unifies diverse quantization formats but suffers from limited hardware support. In this work, we propose a new method that unifies the two paradigms by leveraging the distinct characteristics of the prefill and decoding phases in LLM inference. \ting{I think we can delete this paragraph. }

\subsection{NPUs on Mobile SoCs}


\begin{figure}[t]
\centering
\includegraphics[width=0.8\linewidth]{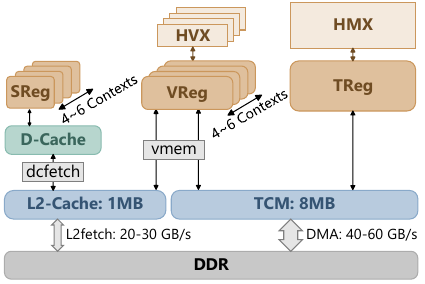} 
\caption{A typical NPU architecture, shown with Snapdragon8 NPU. It integrates the matrix core (HMX), vector cores (HVX), scalar units, and the on-chip memory (TCM). }
\label{fig:npu-arch}
\end{figure}

To support the widely used AI applications on devices, NPUs are designed for utmost performance and efficiency to accelerate neural network inference. These designs emphasize hardware specialization for the most computation-intensive kernels, i.e. dense GEMMs, at certain numerical precisions, while eliminating general-purpose control units that add overheads. Although NPU implementations vary across vendors~\cite{npulist}, such as the ones from Huawei and MediaTek, their architecture shares similar design principles. 

\paragraph{Snapdragon NPU} We use the Qualcomm Snapdragon NPU as a representative example. Snapdragon NPU features top computation performance among the mobile NPUs and relatively accessible SDKs. They are widely used by diverse flagship mobile devices from Microsoft, Samsung, Xiaomi, Honor, OnePlus etc. The Snapdragon X Elite NPU has 45\,TOPS peak performance, much higher compared to the 4.6\,TFLOPS GPU and <1 TOPS CPUs on the same SoC~\cite{xelite}. 

As shown in Fig.~\ref{fig:npu-arch}, the computing units of a Snapdragon NPU consist of one matrix core known as HMX (Hexagon Matrix eXtensions) operating on a $32\times32$ tile, the 4 to 6 vector cores known as HVX (Hexagon Vector eXtension) with a $1\times256$ tile, and the scalar computing units.

For the memory hierarchy, each of the computation units has its own dedicated register file. The vector and scalar registers can keep 4 to 6 contexts, enabling 4 to 6 hardware threads to run concurrently.  Beneath the registers is the large software-managed on-chip memory, named TCM (Tightly Coupled Memory), with a capacity of 8\,MB and a burst transfer width of 2\,KB from TCM to HMX, matching the grand computation width for HMX and HVX. There is also the general L2 cache (1\,MB size and 128\,B access width) shared by the vector and scalar units. Data loading from the off-chip DDR to the TCM is via DMA path which can achieve 40-60\,GB/s, while to the L2 cache can achieve 20-30\,GB/s (detailed measurements in Table~\ref{tab:memory_bandwidth}).  

\paragraph{QNN SDK} The primary SDK for developing on Qualcomm NPUs is the Qualcomm AI Engine, aka the QNN SDK~\cite{qnnsdk}. It is closed source. On-device inference frameworks such as ExecuTorch~\cite{executorch}, ONNX Runtime~\cite{onnxqnn}, llm.npu~\cite{llm.npu}, and PowerServe~\cite{powerserve}, leverage QNN as their backend for NPUs. Developers cannot easily customize low-level kernels. To our knowledge, \sys{} is the first to provide low-bit kernels for diverse quantization algorithms for Qualcomm NPUs.

\begin{figure}[t]
\centering
\includegraphics[width=\linewidth]{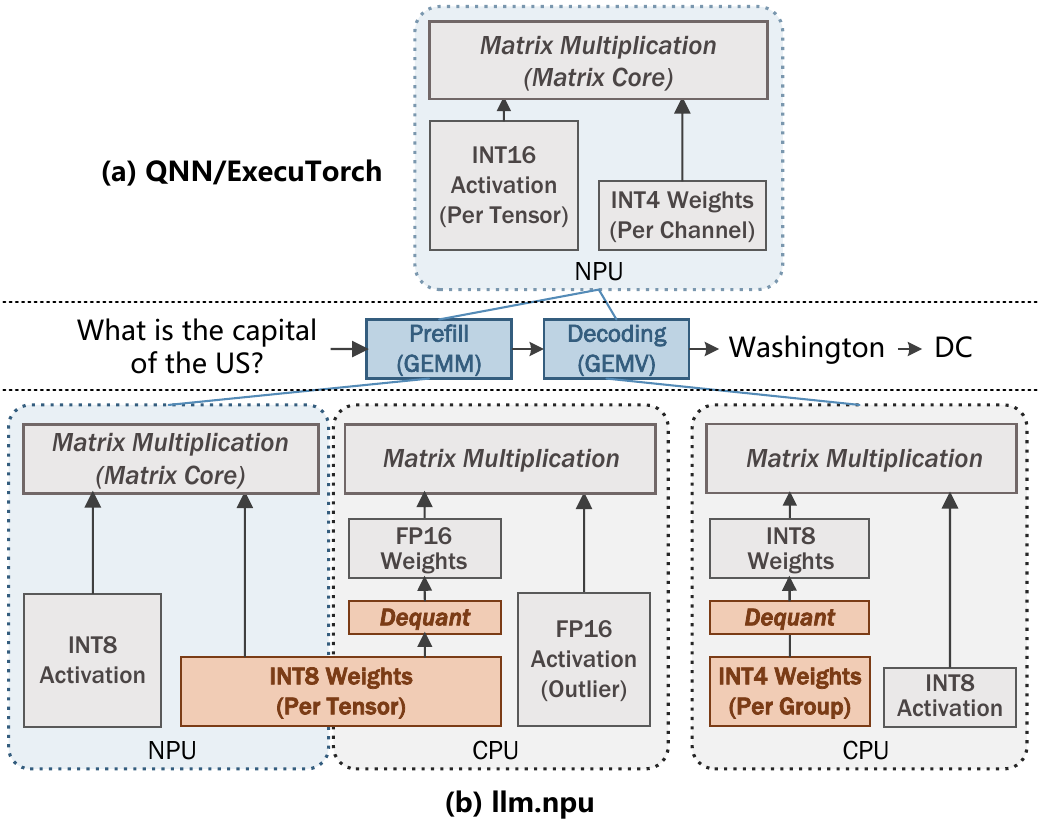} 
\caption{Current practices in leveraging NPUs for low-bit LLM inference, illustrated with SOTA frameworks: Qualcomm QNN from industry and llm.npu from academia. }
\label{fig:existing_practice}
\end{figure}

\section{Challenges of Low-Bit LLMs on NPUs}





As explained in Section~\ref{sec:low-bit-approach}, the main challenge in deploying low-bit LLMs on NPUs still lies in the mismatch between hardware specialization and the diversity of quantization algorithms. As outlined below, each of the current practices has inherent deficiencies that prevent them from fully addressing the challenge.

\begin{figure}[t]
    \centering
    \includegraphics[width=\linewidth]{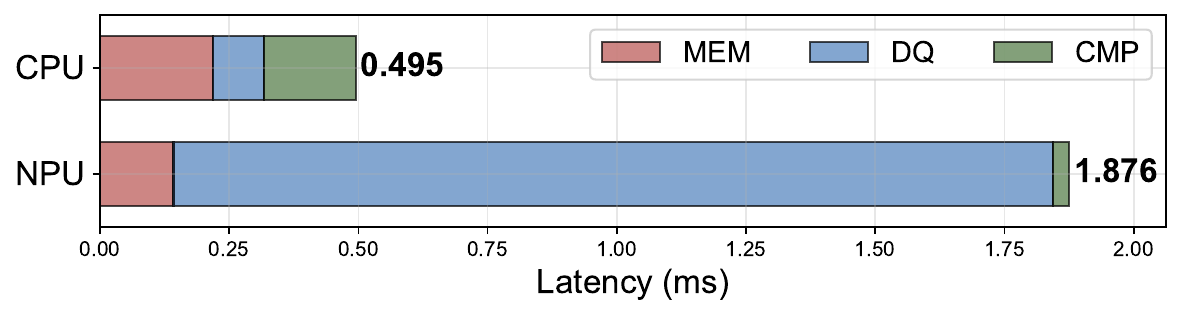}
    \caption{Latency breakdown for a mpGEMV of size $4096 \times 4096 \times 1$, comparing NPU and CPU performance. The total latency is segmented into memory loading (MEM), dequantization (DQ), and computation (CMP).}
    \label{fig:dequant_cost_NPU}
\end{figure}

\paragraph{Slow dequantization.} A straightforward way to bridge this gap is through dequantization, which converts arbitrary quantization formats into the NPU’s supported format. However, as shown in Fig.~\ref{fig:dequant_cost_NPU}, which is a $W_4A_{16}$ mixed-precision GEMV kernel of Llama3 decoding. We can see the kernel runs $3.8\times$ slower on the NPU than the CPU, due to the $10\times$ slower dequantization speed. 
For the prefill stage, which is GEMM-based, the dequantization overhead can be offset by NPU’s superior GEMM performance. 
In contrast, the decoding stage relies on memory-bound GEMV operations, where the powerful matrix core is idle. 
In this setting, the dequantization step emerges as a dominant source of overhead.

\paragraph{Accuracy loss using hardware format} To avoid the dequantization, some frameworks represented by the Qualcomm QNN enforces quantization match the NPU format, shown in Fig.~\ref{fig:existing_practice}(a). 
However, the restricted hardware formats often fail to capture the weight distribution, leading to accuracy loss. As will be presented in Sec.~\ref{subsec:accuracy}, using the NPU’s per-channel INT4 weight quantization with per-tensor INT16 activation results in 1.45$\times$ worse perplexity.

\paragraph{Energy overhead of NPU for prefill and CPU for decoding} To exploit the NPU’s GEMM performance without sacrificing accuracy, some frameworks represented by llm.npu only run prefill on the NPU while leaving decoding and outlier computation on the CPU. As shown in Fig.~\ref{fig:existing_practice}(b), during prefill, NPU uses INT8-quantized weights and executes the INT8 GEMM on the matrix core for the best performance. During decoding, the CPU uses INT4-quantized weights, dequantizes them to INT8, and executes INT8 GEMV using SIMD units. While this hybrid approach achieves latency gain, it increases energy consumption by up to 3.4$\times$.

Overall, existing methods fail to achieve simultaneous improvements in latency, energy efficiency, and accuracy when deploying low-bit LLMs on edge NPUs.  

\section{Design}

To enable efficient prefill and decoding on an NPU, we propose a two-pronged approach. For the prefill stage, we employ a dequantization-based GEMM to maximize the use of the NPU's powerful matrix core and maintain flexibility with various quantization algorithms. For the decoding stage, we use a LUT-based GEMV on the less powerful vector core, which eliminates the overhead of dequantization and reduces computational demands.

These distinct computational patterns require weights to be permuted and reshuffled for ideal memory access pattern during execution, and lead to conflicting data layout requirements in three ways: (1) different bit-packing methods (bit-parallel vs. bit-serial), (2) different precision required and (3) mismatched tiling strategies dictated by the computation and register constraints. A naive approach would require two separate model weights, doubling the memory footprint.

\begin{figure}[t]
    \centering
    \includegraphics[width=\linewidth]{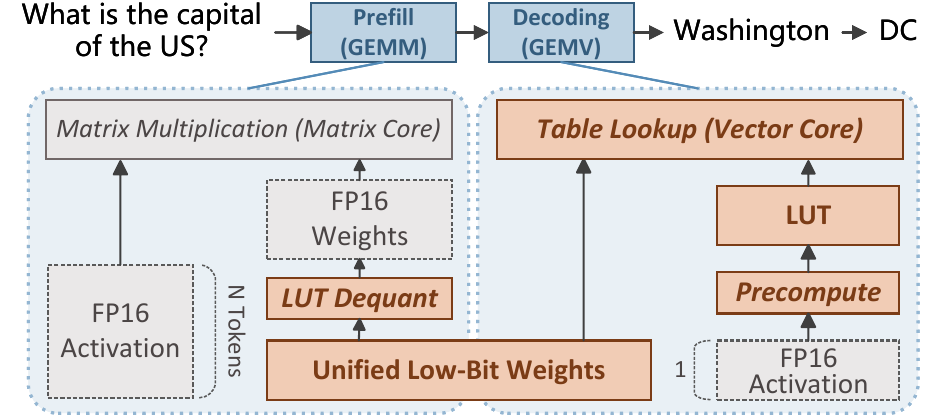}
    \caption{\sys{} system. It runs both prefill and decoding on NPU through table lookup. Only one copy of model weights.}
    \label{fig:system}
\end{figure}

To resolve, our key idea is to co-design a single, unified weight layout and tiling strategy that efficiently serves both computation patterns. The overview of the system design is presented in Fig.~\ref{fig:system}.


\subsection{Unified Low-Bit Weights for Prefill }

\paragraph{Unified Layout through Fused LUT-based Dequantization}

Due to the generative, token-by-token nature of the decoding stage, performing any layout transformation introduces significant overhead. Therefore we prioritize the layout required for decoding by using bit-serial packing. Consequently, the bit-repacking is performed during prefill.

However, designing an efficient on-the-fly dequantization process presents two primary challenges: (1) a data layout mismatch exists between the bit-serial nature of LUT-decoding and the bit-parallel required for dequantization, requiring inefficient bit-shuffling operations and (2) NPUs are primarily designed for low power and fast integer operations, feature much slower floating-point conversion and computation, making a naive dequantization implementation the major bottleneck.

\begin{figure}[t]
    \centering
    \includegraphics[width=\linewidth]{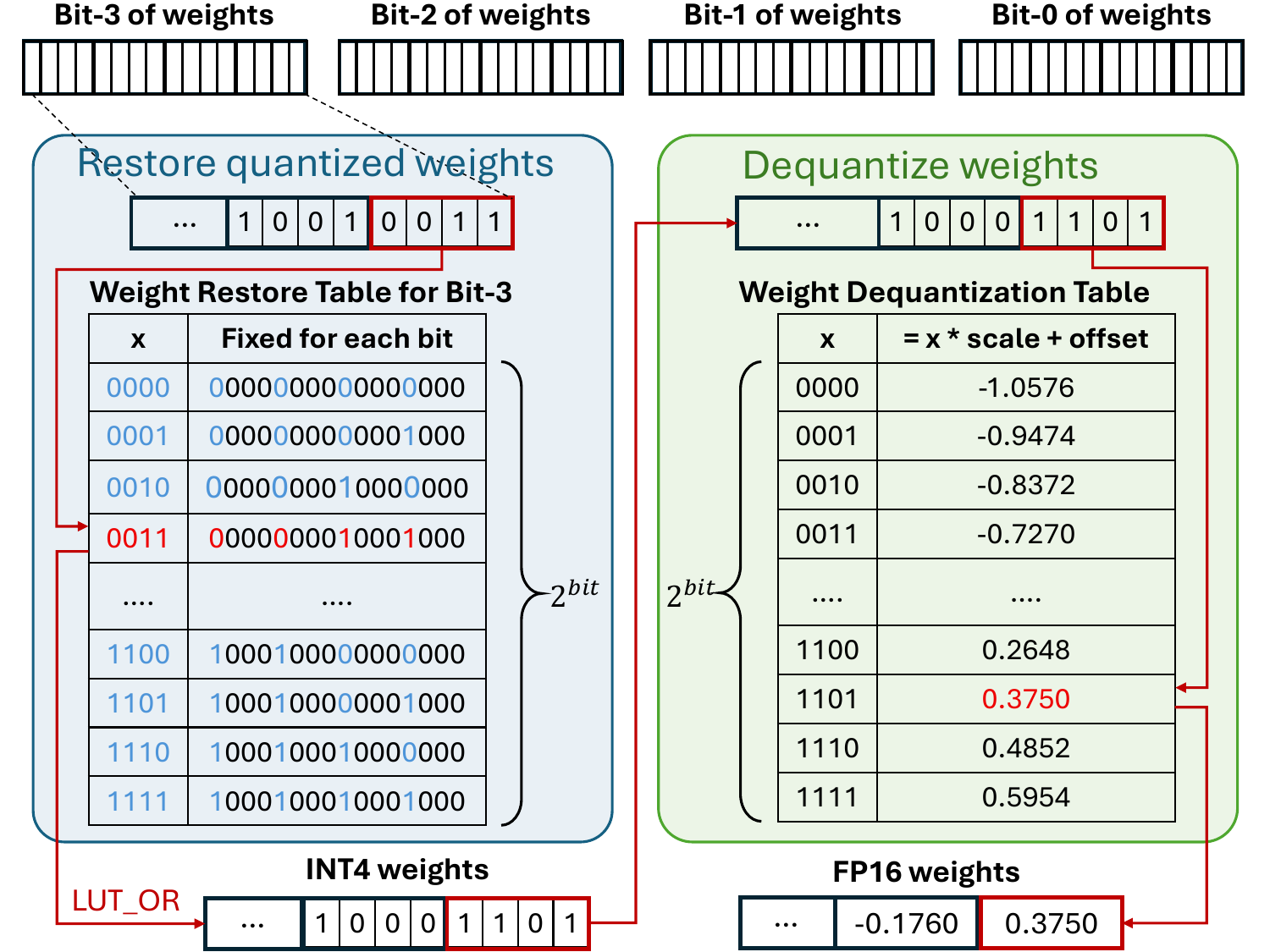}
    \caption{Fused two-level LUT dequantization. Take 4-bit quantized weights as an example.}
    \label{fig:lut-dequant}
\end{figure}

To overcome the inefficiency of bit-shuffling and floating-point math operations, we introduce a highly efficient dequantization method that consolidates multiple steps into two distinct LUT operations, as illustrated in Fig.~\ref{fig:lut-dequant}.

\textit{Bit repacking.} A naive approach to repacking data from a bit-serial to a bit-parallel layout involves a slow sequence of bitwise operations (e.g., \code{AND}, \code{OR}, \code{SHIFT}) that handle only one bit at a time. Our method accelerates this process by treating a group of packed bits (for instance, the $i$-th bit from four different weights) as a direct index into a repacking LUT. The table entry at that index contains the pre-calculated, correctly arranged bit-parallel representation. For example, the packed value \code{0b0011}, representing the most significant bit (MSB) of four INT4 weights, is used to fetch the entry \code{0b0000 0000 1000 1000}. This single lookup instantly places the MSB for each of the four weights into its correct final position. After this is done for all bit positions, the results are combined to reconstruct the original quantized weights. By using a 4-bit LUT index, we replace twelve separate bit manipulation operations (four sets of \code{SHIFT+AND+SHIFT}) with just one LUT lookup, achieving a $12\times$ reduction in operations for this step.

\textit{Integer-to-float conversion.} Instead of performing costly runtime integer-to-float operations, we exploit the small number of unique values in low-bit formats (e.g., 16 for INT4). We pre-calculate the FP16 representation for each of the 16 possible integer values and store them in a conversion LUT.

\textit{Applying scales and zero-points.} We fuse the affine transformation into a LUT by pre-applying the quantization scales and zero-points to its entries. This "bakes" the transformation directly into the table, which can be shared in the quantization block. For INT2 values and typical quantization block sizes of 64 and 128, this approach reduces the needed floating operations to 4, which is 1/16 and 1/32 of the original respectively. This method effectively removes the main computational bottleneck of the dequantization procedure.

Even with the bit-packing resolved, the mismatched tiling requirements still create memory access overhead during both prefill and decoding. To understand and address the tiling constraints, we first abstract a unified concurrency model that applies to both the vector and matrix cores.

\paragraph{NPU concurrency hierarchy}

We analyze the typical NPU's memory hierarchy and SIMD programming model, and summarize the concurrency hierarchy to three levels: (1) \textit{pipeline-level}, where DMA, vector cores, and matrix core execute concurrently, (2) \textit{thread-Level}, where data is processed in parallel by multiple threads on vector cores to compute or prepare for the matrix core, and (3) \textit{SIMD-level}, where data is loaded into vector or matrix registers for SIMD computation.

This hierarchical structure demands a careful tiling design to efficiently fit data into on-chip memory.

\paragraph{Unified tiling strategy}

\begin{figure}[t]
    \centering
    \includegraphics[width=\linewidth]{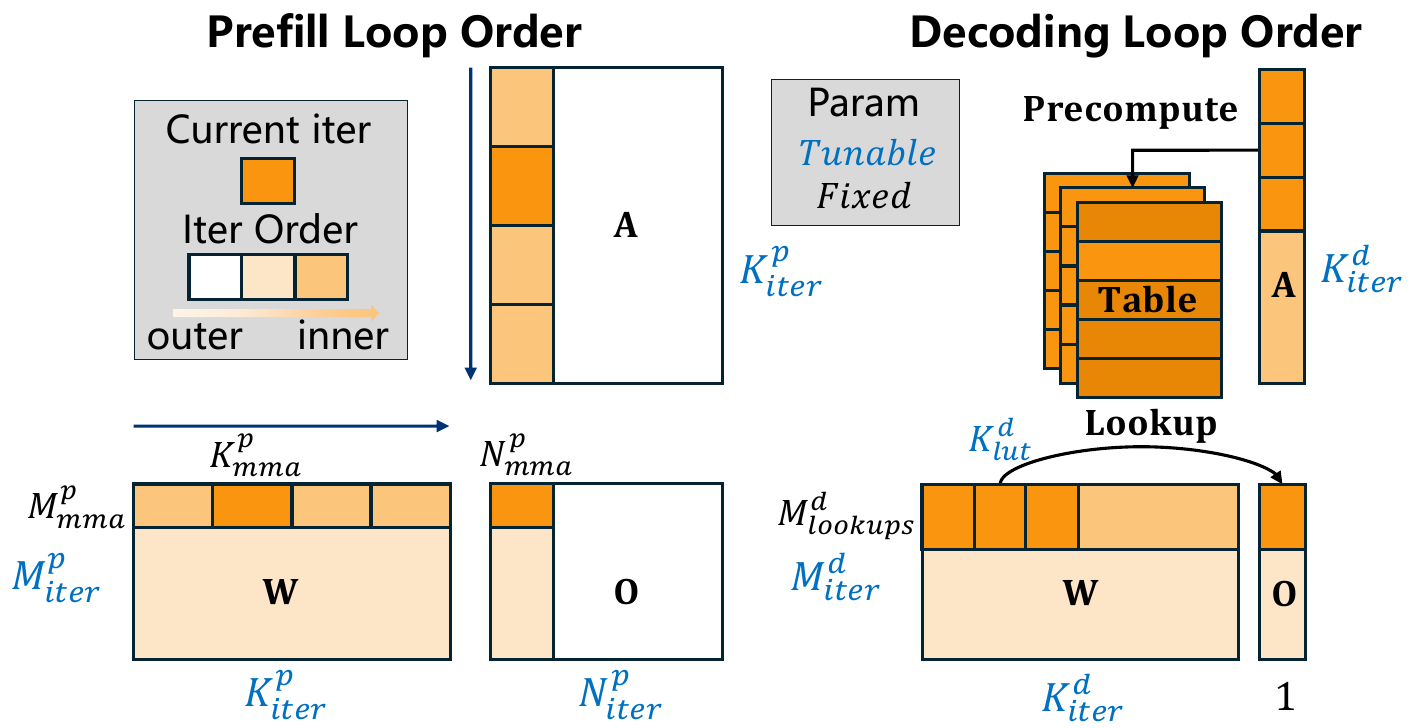}
    \caption{Thread-level tiling and loop orders.}
    \label{fig:tile-loop}
\end{figure}

The optimal tiling for dequantization-based prefill differs significantly from that of LUT-based decoding. As illustrated in Fig.~\ref{fig:tile-loop}, for a matrix multiplication of weights $(M, K)$ and activations $(N, K)$, the thread-level tiling and loop orders are:

\begin{itemize}
    \item \textbf{Prefill}: $(N^{p}_{iter}, M^{p}_{iter}, K^{p}_{iter}, N^{p}_{mma}, K^{p}_{mma}, M^{p}_{mma})$, where $\{N, M, K\}^{p}_{iter}$ are tunable iteration counts, $\{N, M, K\}^{p}_{mma}$ are fixed dimensions of the matrix core's MMA instruction.
    \item \textbf{Decoding}: $(K^{d}_{iter}, M^d_{iter}, K^{d}_{lut}, M^{d}_{lookups})$, where $K^{d}_{iter}$, $M^{d}_{iter}$, and $K^{d}_{lut}$ (number of lookup tables in registers) are tunable, while $M^{d}_{lookups}$ is a fixed number determined by the vector length.
\end{itemize}

The primary challenge is creating a single, pre-permuted weight layout that serves both tiling schemes. This is critical because weights must be fetched into on-chip memory via DMA in a contiguous block. To find a valid unified tiling, we define a search space following the constraints as:
\begin{align}
    K^{d}_{lut} &< \text{N\_REG} \label{eqn:1} \\
    M^{p}_{iter} \cdot M^{p}_{mma} &= M^d_{iter} \cdot M^{d}_{lookups} \label{eqn:2} \\
    K^{p}_{iter} \cdot K^{p}_{mma} &= K^{d}_{iter} \cdot K^{d}_{lut} \label{eqn:3} \\
    \text{N\_STAGE} \cdot \text{N\_THREAD} &\cdot S_{tile} < S_{on\_chip\_mem} \label{eqn:4} \\
    \text{where } S_{tile} = M^{p}_{iter} \cdot &M^{p}_{mma} \cdot K^{p}_{iter} \cdot K^{p}_{mma} \nonumber
\end{align}
To explain, the tunable parameters are governed by register and TCM capacity. $K^{d}_{lut}$ must be less than the number of vector registers (Eqn.~\ref{eqn:1}). A unified permutation requires the tile dimensions to match (Eqn.~\ref{eqn:2},~\ref{eqn:3}), and the total memory footprint of all pipeline stages and parallel threads must not exceed the on-chip memory size (Eqn.~\ref{eqn:4}).

To direct the search, we analyze the characteristics of the NPU, and provide the following heuristics:

\begin{itemize}
    \item Maximize $K^{d}_{lut}$ to keep more lookup tables in registers, reducing intermediate write-backs brought by the LUT-based computing pattern as described in Sec.~\ref{subsec:decoding}.
    \item Maximize $M^{d}_{iter}$ to increase the data reuse of cached lookup tables.
    \item Maximize $K^{p}_{iter}$ to improve the throughput of the matrix core.
\end{itemize}

Finally, while the thread-level data layout must be contiguous for DMA, the mismatched axis order at the finer-grained SIMD level is acceptable, as non-contiguous loads from TCM to registers do not incur a performance penalty.

\subsection{Pipelining to Hide Dequantization Overhead}

\label{subsec:prefill}

\begin{figure}[t]
    \centering
    \includegraphics[width=\linewidth]{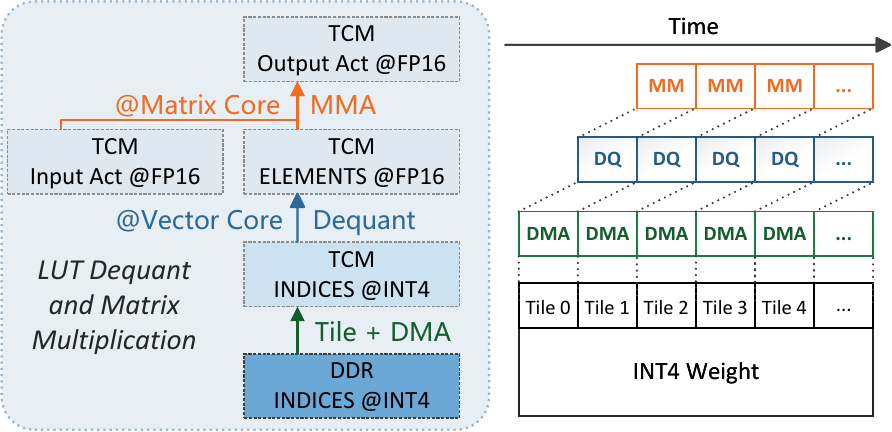}
    \caption{Pipeline of DMA transfer, vector-core dequantization and matrix-core matrix multiplication to hide the latency of memory access and dequantization.}
    \label{fig:pipeline}
\end{figure}

We can further mask the latency of the dequantization by creating a \textit{DMA-Vector-Matrix three-stage pipeline} that overlaps data movement, data preparation and data consumption. As shown in Fig.~\ref{fig:pipeline}, our approach pipelines three key stages:

\begin{itemize}
    \item \textbf{DMA transfer:} The DMA engine streams quantized weights from high-latency DDR to the on-chip TCM. The weights are tiled to fit into TCM.

    \item \textbf{Vector-core dequantization:} Concurrently, the vector core dequantizes the previous tile (already in TCM) into FP16 using our fused LUT technique.

    \item \textbf{Matrix-core matrix multiplication:} Simultaneously, the matrix core performs matrix multiplication on the tile prepared in the prior cycle.
\end{itemize}

This pipelined execution ensures that the memory bus, vector core, and matrix core operate in parallel, maximizing hardware utilization and hiding the dequantization overhead.

\subsection{Mapping of LUT-based Decoding on NPU}

\label{subsec:decoding}

While dequantization offers flexibility, it creates significant computational overhead and under-utilizes matrix cores during decoding. LUT based method offers a more elegant solution. By treating different data types as table indices and entries, it eliminate the need for dequantization and fully leverage vector cores.

However, unlike traditional dot-product-based methods that vectorize along the input channel ($k$-axis), LUT-based methods vectorize along the output channel ($m$-axis). This alternative approach significantly increases the amount of intermediates that must be managed within each computational tile.

As illustrated in Fig.\ref{fig:lut-gemm}, at the SIMD level, a single operation involves using a vector of indices to perform many parallel lookups into the same LUT. The results from these lookups belong to different output channels and therefore form a vector of partial results, which cannot be immediately aggregated into a single scalar. To handle these large amounts of intermediates, frameworks such as T-MAC align their computation tile size with the quantization block size. This allows all necessary operations—including lookups, partial aggregations, and scaling—to be performed locally within the tile, thereby minimizing costly data movement between the processing unit and memory.

However, this strategy is suboptimal for modern NPUs, which typically have wide vector length (e.g., 1024-bit). Fully exploiting this hardware requires a tile size far larger than a typical quantization block. For example, to optimally use 16 registers reserved for LUTs on Qualcomm NPUs, the tile size on the $k$-axis needs to be 256, which is much larger than the common quantization block sizes of 64 or 128.

To bridge this gap, \sys employs a two-level tiling strategy. The smaller, inner tile aligns with the quantization block for efficient low-precision aggregation. The outer tile is sized to process as many LUTs as can be stored in registers, allowing extensive aggregation of floating-point results before writing to memory. While this maximizes data reuse, it introduces a new bottleneck: the outer tile requires more floating-point accumulators than the hardware register file can hold. This forces the compiler to spill excess registers to the slow L2 cache, severely degrading performance.

To resolve this, we introduce a software-managed register-spill buffer located in the fast, on-chip TCM. This buffer acts as a programmer-controlled extension of the register file, efficiently managing the storage of intermediate accumulators without resorting to high-latency cache. 

\begin{figure}[t]
    \centering
    \includegraphics[width=\linewidth]{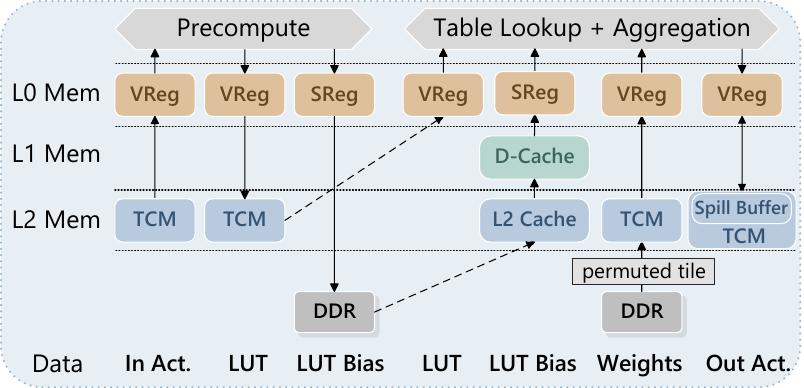}
    \caption{Map LUT decoding memory hierarchy onto NPU.}
    \label{fig:lut-mem}
\end{figure}

Fig.~\ref{fig:lut-mem} depicts the mapping of the LUT decoding onto the NPU memory hierarchy. TCM stores input activations and most intermediates, L2 cache is reserved for scalar values, and the TCM-based spill buffer manages aggregation results.

\section{Implementation}

\sys consists of 7,100 lines of C++ and Python code and is built upon the open-source ExecuTorch library. Key contributions of our implementation include highly optimized LUT-decoding NPU kernels, extended support for custom QNN operator packages in ExecuTorch, and the implementation of more LLMs and quantization schemes.

\paragraph{Graph optimization}
LLMs contain layers where the same activation feeds multiple GEMM operations (e.g., Q/K/V projections in attention, up/gate projections in MLPs). One of the practices is to fuse these GEMMs into a larger GEMM, but it is not viable on the NPU, whose limited on-chip memory TCM (8MB) favors splitting them rather than merging them.

\begin{figure}[t]
    \centering
    \includegraphics[width=\linewidth]{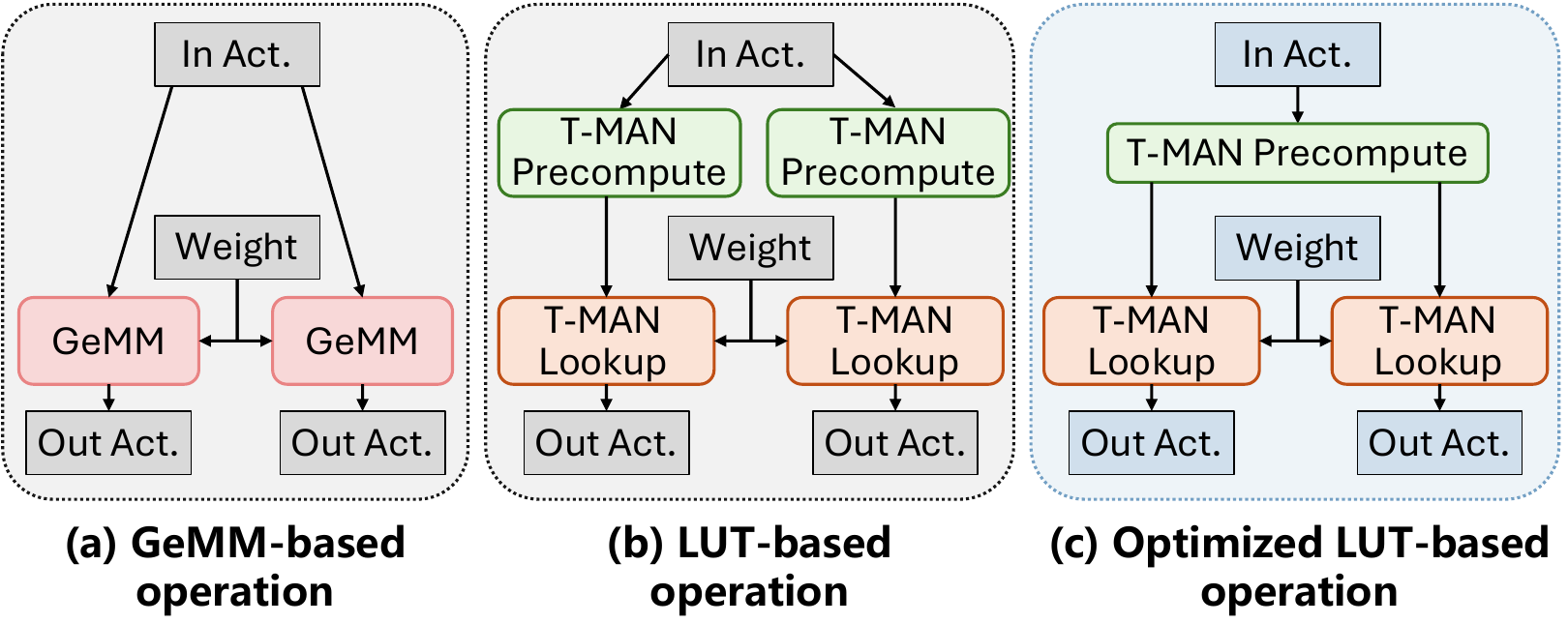}
    \caption{Graph Optimization.}
    \label{fig:graph-optimization}
\end{figure}

For our LUT-based kernels, this pattern leads to redundant computation and excessive memory consumption when the small kernels are scheduled to execute in parallel. We address this by unfusing our kernel into two components: a precomputation kernel and a table look-up kernel. A graph optimization pass then detects shared input patterns and prunes redundant precomputation kernels. As shown in Figure~\ref{fig:graph-optimization}, this ensures that multiple table look-up kernels reuse the output from a single precomputation kernel, saving both cycles and memory.

\paragraph{Efficient table look-up by VLUT}

To perform efficient table look-ups, \sys leverages the VLUT SIMD instruction from HVX. VLUT has two variants: VLUT16, which uses a table of 16 entries with 16 bits per entry, and VLUT32, which uses a table of 32 entries with 8 bits per entry. Both instructions take 8-bit values as indices into the table.

To determine the optimal variant, we analyzed their computational throughput in terms of equivalent Multiply-Adds (MADDs), as detailed in Table~\ref{tab:vlut_analysis}. It shows that VLUT16 achieves higher throughput for both 8-bit and 16-bit activations. We thus select VLUT16 for our implementation.


\begin{table}[t]
    \centering
    \resizebox{\linewidth}{!}{
    \begin{tabular}{ccccc}
         \toprule
         & \textbf{Bitwidth} & \textbf{CPI} & \# \textbf{Look-ups} & \# \textbf{Equiv. MADDs} \\
         \midrule
         \multirow{2}{*}{\textbf{VLUT16}} & 8            & 0.5 & 256        & 1024  \\
                                 & 16           & 0.5 & 128        & 512   \\
         \multirow{2}{*}{\textbf{VLUT32}} & 8            & 0.5 & 128        & 640   \\
                                 & 16           & 0.5 & 64         & 320   \\
         \bottomrule
    \end{tabular}
    }
    \caption{Comparison of VLUT16 and VLUT32 throughput.}
    \label{tab:vlut_analysis}
\end{table}


\paragraph{Asynchronous DMA}

Efficiently transferring data from main memory (DDR) to the processor is critical for performance, especially for the decoding stage. To identify the optimal method, we benchmarked available approaches provided by the NPU on OnePlus 12 Pro. The results in Table \ref{tab:memory_bandwidth} clearly show that DMA achieves the highest and most consistent bandwidth (59 GB/s). The "Vectorized Load" method causes large pipeline stalls due to high memory latency, resulting in the lowest bandwidth.

\begin{table}[t]
    \centering
    \resizebox{\linewidth}{!}{\begin{tabular}{lcc}
         \toprule
         \multirow{2}{*}{\textbf{Method}} & \textbf{Bandwidth} & \textbf{Bandwidth} \\
                                          & (HVX\_THREADS=1)    & (HVX\_THREADS=4)    \\
         \midrule
         \textbf{Vectorized Load} & 5~GB/s & 20~GB/s \\
         \textbf{L2fetch} & 26~GB/s & 32~GB/s \\
         \textbf{DMA} & 59~GB/s & 59~GB/s \\
         \bottomrule
    \end{tabular}
    }
    \caption{Memory bandwidth microbenchmark on OnePlus 12. (1) \textbf{Vectorized Load} implicitly caches data in L2 before loading it to a vector register, (2) \textbf{L2 Prefetch} uses the l2fetch instruction to explicitly bring data into the L2 cache, and (3) \textbf{DMA} asynchronously transfers data directly from DDR to the NPU's TCM.}
    \label{tab:memory_bandwidth}
\end{table}

Based on our findings, we use DMA to load model weights and employ software pipelining to hide memory latency. For smaller data chunks and scalar values that should not be placed in vector registers, we use \code{l2fetch}.


\section{Evaluation}

To demonstrate its efficiency, we evaluate \sys using real-world low-bit kernels from a diverse set of LLMs, including Qwen, Llama3, and BitNet. We then compare the end-to-end throughput for both prefill and decoding against leading open-source (llama.cpp~\cite{llama.cpp}, llm.npu~\cite{llm.npu}, T-MAC~\cite{t-mac}, bitnet.cpp~\cite{bitnet.cpp}) and closed-source (QNN~\cite{qnnsdk}) frameworks. The primary findings of our analysis are summarized as follows:

\begin{itemize}
    \item Through its efficient LUT-based kernels on the NPU, \sys achieves performance advantage even when compared to framework through vendor-specific hardware acceleration like QNN.
    \item \sys supports popular quantization schemes that are unavailable in existing NPU frameworks, including per-group quantization, which is essential for quantization accuracy, and flexible precision combinations such as 2-bit weights.
    \item By executing all computations exclusively on the power-efficient NPU, \sys substantially reduces power consumption compared to both CPU-only and hybrid CPU-NPU solutions.
\end{itemize}

\subsection{Evaluation Setup}

\paragraph{Hardware}
We evaluate \sys on two smartphones: OnePlus 12, equipped with Qualcomm Snapdragon 8 Gen 3 and 24~GB of RAM, and OnePlus 13T, with Qualcomm Snapdragon 8 Elite and 12~GB of RAM.

\paragraph{Models and Quantization}
Our evaluation uses popular LLMs: Qwen3-8B~\cite{qwen3}, Llama-3.1-8B-Instruct~\cite{grattafiori2024llama}, and BitNet-2B~\cite{wang2023bitnet,ma2025bitnetb1582b4ttechnical}. We adopt quantization methods from the state-of-the-art research literature~\cite{frantar2022gptq,du2024bitdistiller,wang2023bitnet} that have been evaluated by the deep learning community. Specifically, the Qwen and Llama models are quantized to 4-bit ($W_{INT4}$) and 2-bit ($W_{INT2}$) in GPTQ~\cite{frantar2022gptq} format using an asymmetric, per-block scheme with a block size of 64. Each block shares quantization scales and zero points. Their inference is performed with a $W_{\{INT4,INT2\}}A_{\{INT16,FP16\}}$ setup. BitNet is evaluated with its native per-tensor 1.58-bit weights ($W_{INT1.58}$). Following standard practice, we treat its ternary weights as 2-bit for inference, corresponding to a $W_{INT2}A_{INT16}$ scheme.

\paragraph{Baselines}
We compare \sys against the following state-of-the-art methods: (1) \textbf{llama.cpp:} A widely-used open-source framework for CPU inference on local devices, supporting per-group quantization, (2) \textbf{T-MAC:} A state-of-the-art LUT-based kernel library for low-bit CPU inference on local devices, supporting per-group quantization, (3) \textbf{bitnet.cpp:} The official inference framework for BitNet, supporting per-tensor quantization, (4) \textbf{llm.npu:} An open-source framework designed for NPU acceleration, which uses per-tensor quantization and offloads outliers to the CPU and (5) \textbf{QNN:} A proprietary, closed-source framework by Qualcomm. It is highly optimized leveraging internal components like HMX, which are not exposed to developers. It is limited to per-channel and per-tensor quantization.

\paragraph{Kernel-level latency}
To ensure consistency, the kernel shapes are taken from the models under evaluation, and the quantization schemes are matched to the source models.


\paragraph{End-to-End throughput}
We evaluate end-to-end performance by measuring throughput for a prefill phase with a 1024-token input prompt, followed by a generation phase of 128 tokens, using a batch size of one.

\subsection{mpGEMM/mpGEMV Kernel Benchmark}

We evaluate the mpGEMM/mpGEMV kernels in Qwen3-8B, Llama-3.1-8B and BitNet-2B on both devices.

\paragraph{mpGEMV}

\begin{figure*}[t]
    \centering
    \includegraphics[width=\linewidth]{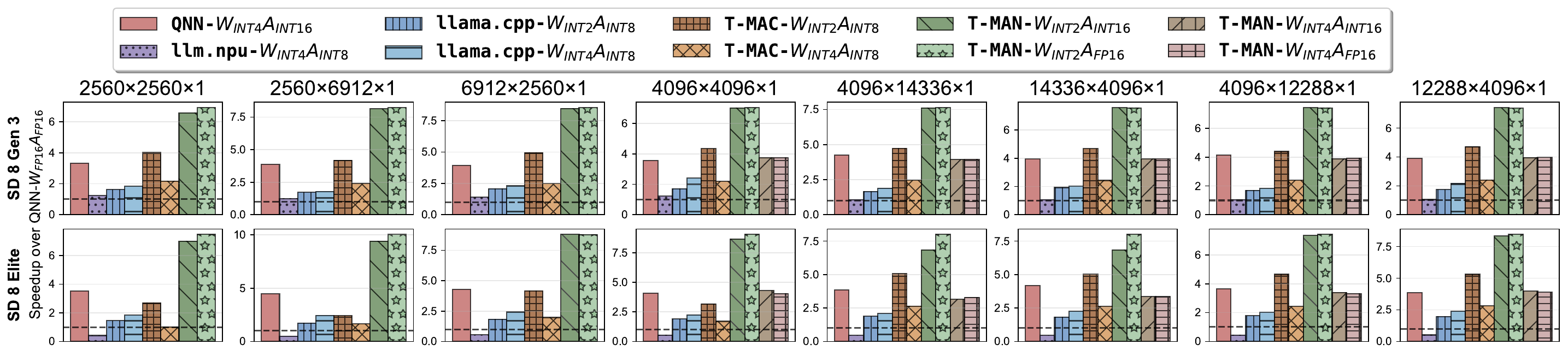}
    \caption{Performance benchmark of the mpGEMV kernel. \sys{}, llama.cpp, and T-MAC use per-block quantization, except for BitNet kernels (shapes \{2560,6912\}×\{2560,6912\}) which use per-tensor. QNN is evaluated with per-channel quantization.}
    \label{fig:eval-gemv}
\end{figure*}

In addition to introducing NPU support for $W_{INT2}$ and per-group quantization, \sys achieves significant speedups as shown in Fig.~\ref{fig:eval-gemv}.

\sys achieves up to 8$\times$ speedup compared to QNN-$W_{FP16}A_{FP16}$. llm.npu fails to accelerate the decoding kernel, as high communication costs from offloading outlier calculations force it to fall back to CPU-only kernels.

\sys is even 1.8–2.5$\times$ faster than QNN for 2-bit kernels and achieves similar performance on 4-bit kernels, despite using a more fine-grained quantization scheme (per-block vs. per-channel). This demonstrates the effectiveness of our software-level optimizations over the framework built upon vendor-specific acceleration.

\paragraph{mpGEMM}
\sys dequantizes the per-tensor quantized weights in BitNet kernels to INT8 and the per-block quantized weights in Qwen3-8B and Llama-3.1-8B-Instruct to FP16. The latency of this dequantization step is effectively optimized by LUT-dequantization and pipelining with the matrix multiplication.

\begin{figure*}[t]
    \centering
    \includegraphics[width=\linewidth]{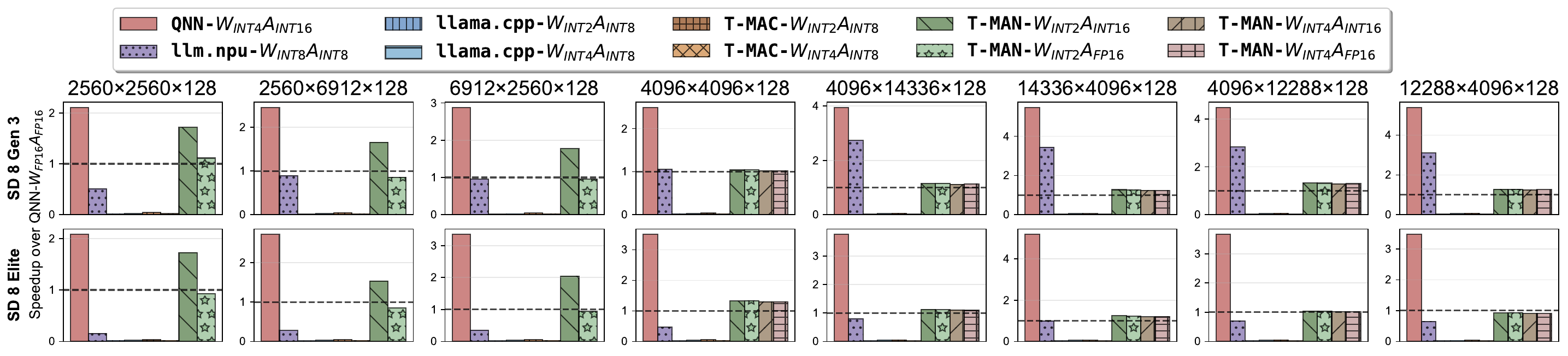}
    \caption{mpGEMM performance comparison. A sequence length of 128 is chosen. It is selected to align with the optimal performance settings for chunked prefill adopted by QNN and llm.npu.}
    \label{fig:eval-gemm}
\end{figure*}

As shown in Fig.~\ref{fig:eval-gemm}, this approach allows \sys to achieve performance comparable to QNN's native $W_{FP16}A_{FP16}$ kernel. Furthermore, \sys is considerably faster than llm.npu on smaller matrix shapes (e.g., 2560×2560×128), as it avoids the NPU-CPU synchronization overhead.
Attributed to the NPU's high throughput (TOPS), \sys delivers a speedup of up to 30$\times$ over CPU-only frameworks like llama.cpp and T-MAC.

\subsection{End-to-End Inference Throughput}

For the decoding phase, \sys demonstrates significant performance advantages, achieving a 1.5-1.8$\times$ speedup over QNN and a substantial 3.1-3.8$\times$ speedup over llm.npu across the tested LLMs. For instance, \sys reaches a notable 49.1~tokens/s on BitNet-2B for Snapdragon 8 Gen 3. llm.npu encounters out-of-memory errors on 8B models on the OnePlus 13T with 12~GB RAM, because it needs to store both INT8 weights for prefill and INT4 weights for decoding.

\begin{figure*}[t]
    \centering
    \includegraphics[width=\linewidth]{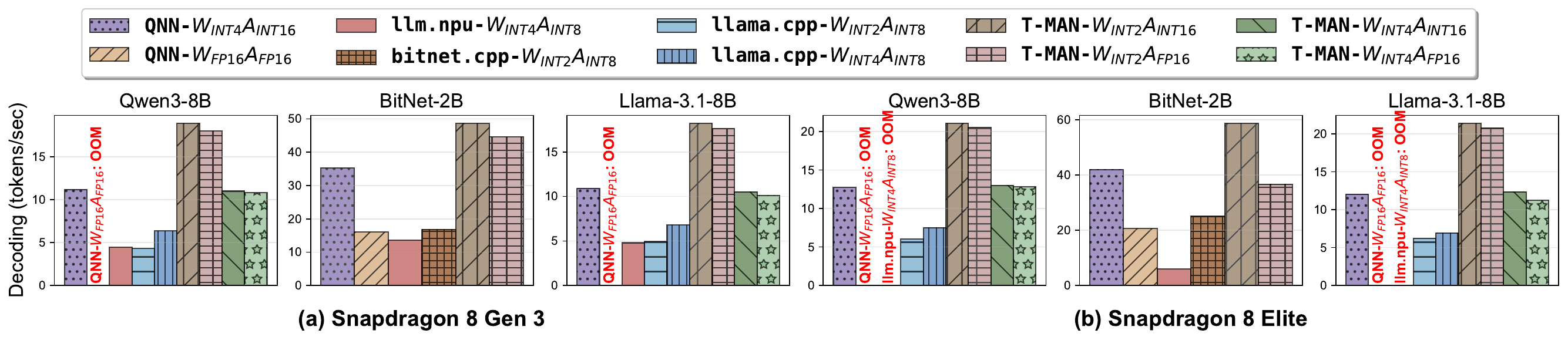}
    \caption{Decoding throughput comparison}
    \hfill
    \label{fig:eval-decoding}
\end{figure*}

In the prefill stage, \sys achieves up to 1.4$\times$ speedup compared to llm.npu, and \sys-$W_{INT2}A_{FP16}$ achieves similar performance compared to QNN-$W_{FP16}A_{FP16}$ on BitNet. When compared to CPU-based solutions like llama.cpp and T-MAC, \sys leverages the high TOPS of NPU to deliver up to 15$\times$ speedup.

\begin{figure*}[t]
    \centering
    \includegraphics[width=\linewidth]{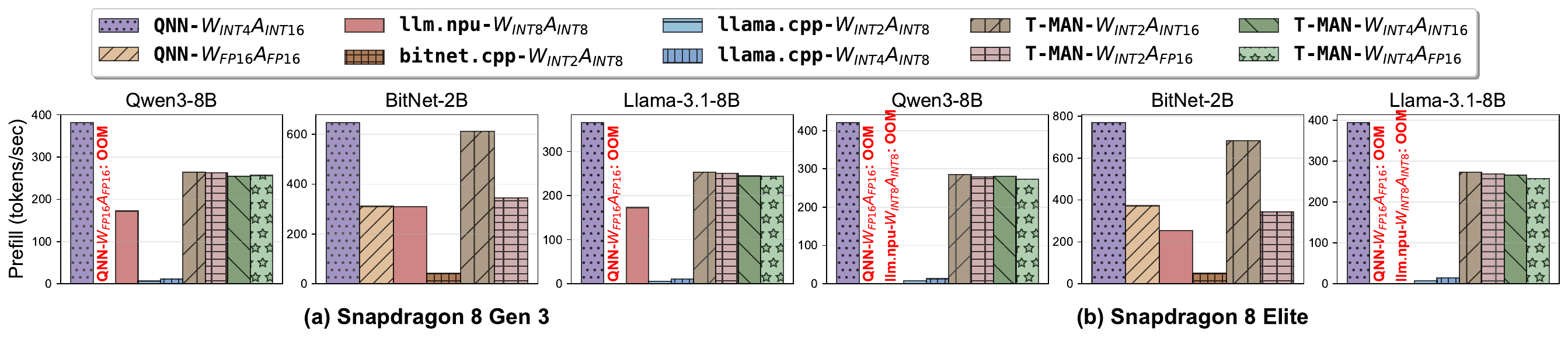}
    \caption{Prefill throughput comparison}
    \label{fig:eval-prefill}
\end{figure*}

This evaluation demonstrates that while supporting more complex and flexible quantization schemes, \sys also maintains comparable performance for prefill and superior performance for decoding.

\subsection{Power and Energy Consumption}
\label{subsec:power}
A key advantage of \sys is its exclusive execution on the power-efficient NPU, eliminating CPU involvement during both prefill and decoding phases. We measure power consumption by reading from \code{/sys/class/power\_supply} at an interval of 100~ms during inference.

\begin{table}[t]
    \centering
    \resizebox{\linewidth}{!}{
    \begin{tabular}{lcccc}
         \toprule
         \multirow{3}{*}{Framework} & \multicolumn{2}{c}{\textbf{Prefill}} & \multicolumn{2}{c}{\textbf{Decoding}} \\
                    & Power & Energy & Power & Energy \\
                    & (W)   & (J/token) & (W) & (J/token) \\
         \midrule
         QNN-$W_{INT4}A_{INT16}$ & 4.96 & \textbf{0.0073} & 4.72 & 0.134 \\
         llm.npu-$W_{INT8}A_{INT8}$ & 8.89 & 0.0269 & - & - \\
         llm.npu-$W_{INT4}A_{INT8}$ & - & - & 8.31 & 0.612 \\
         bitnet.cpp-$W_{INT2}A_{INT8}$ & 8.22 & 0.196 & 8.22 & 0.490 \\
         \sys-$W_{INT2}A_{INT16}$ & 5.01 & \textbf{0.0080} & 4.91 & \textbf{0.101} \\
         \bottomrule
    \end{tabular}
    }
    \caption{Comparison of power and energy efficiency of BitNet-2B on Snapdragon 8 Gen 3.}
    \label{tab:power_and_energy}
\end{table}

Compared to the CPU-only bitnet.cpp, \sys reduces average power consumption by 40\%. When combined with the significant performance improvements, this translates to energy consumption reductions of 24.5$\times$ for prefill and 4.9$\times$ for decoding phases. Compared to the hybrid NPU-CPU llm.npu, \sys reduces power consumption by 45\% for prefill. The unified NPU execution eliminates the power overhead of maintaining active CPU cores for outlier computations, resulting in overall energy savings of 71\% for prefill and 84\% for decoding. Although both \sys and QNN utilize NPU-only computation, \sys still achieves 25\% energy reduction for decoding due to inference speedup.

\subsection{Ablation Study}
We conduct an ablation study to quantify the benefits of: (1) the efficient LUT-based dequantization kernel, and (b) the pipelined execution model.

\subsubsection{Efficiency of the LUT-dequant kernel}
\begin{figure}[t]
    \centering
    \includegraphics[width=\linewidth]{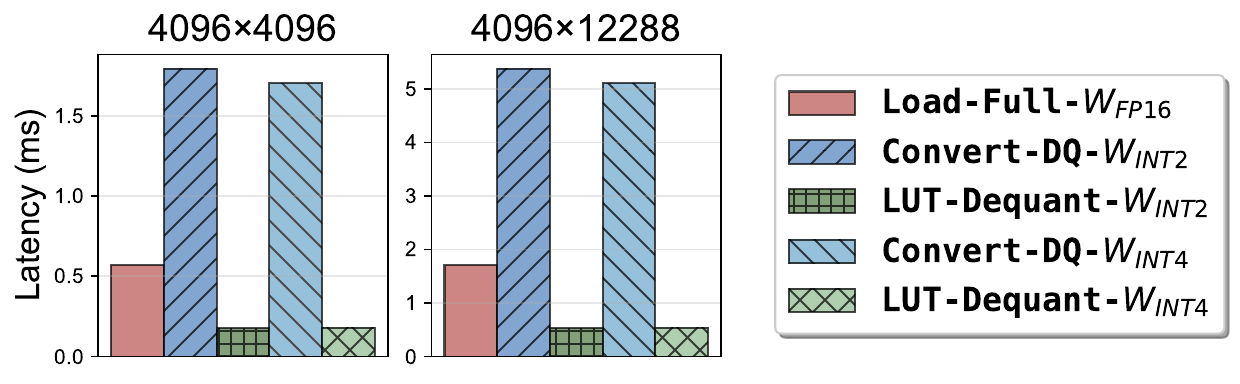}
    \caption{Latency comparison of different methods to prepare full-precision weights on Snapdragon 8 Gen 3.}
    \label{fig:abl-dequant}
\end{figure}

We benchmark our LUT-based dequantization kernel against two baselines: (a) \textit{LoadFull}, which loads pre-converted full-precision weights directly from memory, and (b) \textit{ConvertDQ}, which uses the NPU's standard floating-point operations for dequantization. As shown in Fig.~\ref{fig:abl-dequant}, our kernel achieves a 10.2$\times$ speedup over \textit{ConvertDQ} by avoiding inefficient float conversion instructions. It also outperforms \textit{LoadFull} by 4.9$\times$. This latter gain is attributed to reduced pressure on main DDR memory, as our method loads compact quantized weights and stores the full-precision results in on-chip memory.

\subsubsection{Gain from pipelined execution}
\begin{figure}[t]
    \centering
    \includegraphics[width=\linewidth]{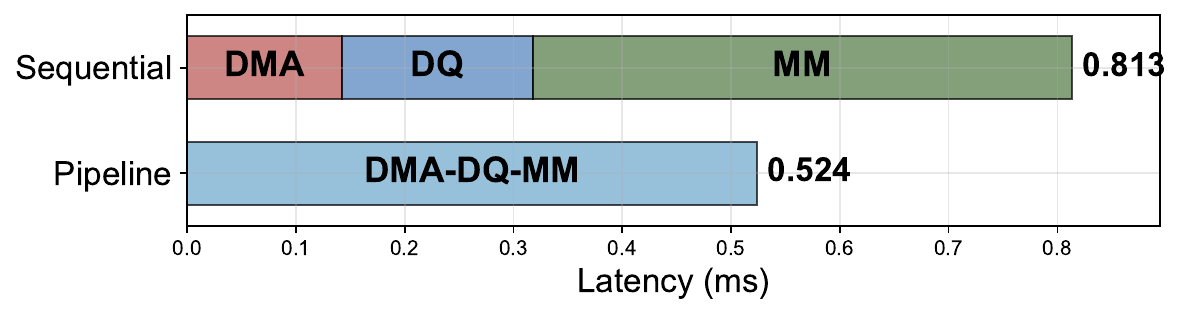}
    \caption{Comparison of sequential vs. pipelined execution for a $4096 \times 4096 \times 128$ $W_{INT4}$ GEMM on Snapdragon 8 Gen3.}
    \label{fig:abl-pipeline}
\end{figure}

Next, we evaluate the performance gain from our three-stage pipeline: (1) \textbf{DMA} loads quantized weights, (2) \textbf{DQ} dequantizes them into on-chip memory, and (3) \textbf{MM} performs the matrix multiplication. As illustrated in Fig.~\ref{fig:abl-pipeline}, pipelined execution achieves a 1.5$\times$ speedup over a sequential approach. The overhead introduced by \sys{} is only 10\% over the matmul stage alone, confirming that our pipeline effectively hides the latency of weight loading and dequantization.

\subsection{Accuracy Benchmark}
\label{subsec:accuracy}
A key advantage of \sys is its support for per-block quantization schemes. We compare its perplexity (PPL) on the WikiText2 dataset against QNN, which is limited to per-tensor and per-channel quantization. Table \ref{tab:acc} shows that \sys not only delivers better throughput but also achieves significantly lower perplexity. Even when using a lower bit-width, \sys achieves perplexity reduction of 48.2\% and 31.7\% for Qwen3-8B and Llama-3.1-8B-Instruct respectively.

\begin{table}[t]
    \centering
    \resizebox{\linewidth}{!}{
    \begin{tabular}{lcc}
         \toprule
         \multirow{2}{*}{Framework} & Llama-3.1-8B-Instruct & Qwen3-8B \\
                                    & PPL $\downarrow$       & PPL $\downarrow$ \\
         \midrule
         QNN-$W_{INT4}A_{INT16}$ & 18.62 & 25.37 \\
         \sys-$W_{INT2}A_{INT16}$ & \textbf{12.81} & \textbf{13.14}\\
         \bottomrule
    \end{tabular}
    }
    \caption{Perplexity on WikiText2. \sys with per-block $W_{INT2}$ outperforms per-channel $W_{INT4}$ used by QNN. \ting{Add MSE of GPTQ 2bit vs ours.}}
    \label{tab:acc}
\end{table}

\section{Discussion and Limitations}

\paragraph{Potential for hardware-accelerated mpGEMM}
While the Qualcomm NPU lacks native 2-bit support, its existing 4-bit matrix multiplication kernels offer a viable path to accelerate the prefill stage of per-tensor quantized BitNet. Our proposed LUT-dequantization method can be applied to dequantize 2-bit weights to 4-bit on-the-fly. However, the implementation of such a custom kernel is currently infeasible, as the hardware instructions for low-bit matrix operations are not yet exposed to developers. The alternative approach, leveraging high-level 4-bit operations via the QNN framework, is constrained by its reliance on a black-box offline weight packing process.

\paragraph{Long-context attention}
Our evaluations are conducted on a context length of 1024 due to the absence of an efficient attention kernel implementation for longer sequences on the NPU. The end-to-end speedup of \sys{} is also bottlenecked by the attention. While the optimization of attention was outside the scope of this paper, it represents a critical area for future work. Such efforts will become more practical as the NPU's matrix core architecture becomes more accessible to developers.

\paragraph{Different NPU hardware}
Although \sys{} was developed and tested on Qualcomm NPUs, its underlying principles are broadly applicable to other NPUs, most of which feature a similar architecture of vector and matrix cores and limited FLOPs. A primary limitation, however, is hardware programmability. Certain platforms, such as Apple NPUs, are closed systems accessible only through high-level APIs, precluding the implementation of custom kernels. Adapting our design to other programmable hardware, like the Ascend NPU~\cite{ascend}, requires further investigation on how to efficiently implement the lookup operation and remains an avenue for future exploration.

\section{Related Works}

\paragraph{LUT for low-bit inference}
An emerging optimization for low-bit models is using LUT to eliminate dequantization and replace expensive multiplications with table lookups. This approach has been applied across hardware, starting with CNNs on CPUs (DeepGEMM~\cite{ganji2023deepgemm}). For LLMs on GPUs, methods like LUT-GEMM~\cite{park2023lutgemm} and FLUTE~\cite{guo2025fastmatrixmultiplicationslookup} developed custom kernels to operate directly on quantized weights, with LUT Tensor Core~\cite{luttensorcore} proposing a software-hardware co-design for further efficiency. T-MAC~\cite{t-mac} pioneers an efficient LUT-based method on CPUs to achieve linear speedup with bit-width reduction.

\paragraph{NPU for LLM inference} Recent research has increasingly focused on leveraging NPUs to address bottlenecks in LLM inference, often by creating heterogeneous systems. For instance, Hybe~\cite{hybe} proposes a GPU-NPU hybrid system, assigning the prefill stage to GPUs and the decoding stage to NPU. The NPU adopts a lightweight architecture specifically designed for decoding. For on-device inference, llm.npu~\cite{llm.npu} presents a hybrid NPU-CPU system where the NPU handles most of the prefill computation while the CPU manages outlier calculations in parallel. Another approach combines NPUs with PIM. NeuPIMs~\cite{neupims} and IANUS~\cite{ianus} exemplify this by using NPUs for GEMM and PIM for GEMV. NeuPIMs focuses on enabling concurrent execution, while IANUS introduces a unified memory system to minimize data movement. Beyond architecture, V10~\cite{v10} improves NPU utilization in cloud inference through a hardware-assisted multi-tenancy framework that enables fine-grained resource sharing and scheduling.

\paragraph{On-device optimization of LLM}

Beyond model quantization and hardware acceleration, algorithmic optimizations are crucial for deploying LLMs on-device. Knowledge distillation is used to create smaller student models by transferring knowledge from larger teacher models~\cite{hsieh2023distillingstepbystepoutperforminglarger}, and can also be applied QAT to mitigate quantization errors~\cite{du2024bitdistiller}. To accelerate inference, speculative decoding breaks the sequential nature of token generation. This technique uses a small draft model to propose candidate tokens that are verified in parallel by the main model. Variations on this approach, such as adding extra decoding heads~\cite{cai2024medusa} or using lookahead strategies~\cite{fu2024break}, further enhance decoding speed and eliminate the need for the draft model.

\ting{Our method is lossless. Other ExecuTorch modules have perf loss.}

\section{Conclusion} 
 This paper realized the first end-to-end LLM inference on the NPUs, achieved the SOTA latency and energy among current on-device inference systems. It is based on the key idea to use table lookups to subsume operations not well supported by the highly specialized hardware. The $1.4\times$ and $3.1\times$ speedup for prefill and decoding, and the energy and memory saving demonstrate the effectiveness of \sys.

\bibliographystyle{ACM-Reference-Format}
\bibliography{references}

\end{document}